\documentclass[espf]{aa}
\usepackage{graphicx}
\usepackage{epsfig}
\usepackage{supertabular}

\titlerunning{A Virtual catalog of calibrators: I The bright objects case }

\begin{document}
\title{\emph{SearchCal}: a Virtual Observatory tool for searching
calibrators in optical long baseline interferometry}

\subtitle{I: The bright object case}

\author{D. Bonneau \inst{1}
                                \and
                J.-M. Clausse \inst{1}
                \and
                X. Delfosse \inst{2}
                \and
                D. Mourard \inst{1}
                \and
                S. Cetre\inst{2}
                \and
                A. Chelli \inst{2}
                \and
                P. Cruzal\`ebes \inst{1}
                \and
                G. Duvert \inst{2}
                \and
                G. Zins \inst{2}
                }
\institute{
                    Observatoire de la C\^ote d'Azur, Dpt. Gemini, UMR
                    6203, F-06130 Grasse, France
                    \and
                    Laboratoire d'Astrophysique de l'Observatoire de
                    Grenoble, UMR 5571, F-38041 Grenoble, France
                    }

\date{Received ... ; accepted ...}

  \abstract
   {In long baseline interferometry, the raw fringe contrast
must be calibrated to obtain the true visibility and then
those observables that can be interpreted in terms of astrophysical
parameters. The selection of suitable calibration stars is crucial
for obtaining the ultimate precision of interferometric instruments
like the VLTI. Potential calibrators must have spectro-photometric
properties and a sky location close to those of the scientific
target. }
   {We have developed software (\emph{SearchCal}) that
builds an evolutive catalog of stars suitable as calibrators
within any given user-defined angular distance and magnitude around the
scientific target. We present the first version of
\emph{SearchCal} dedicated to the bright-object case
(V$\leq$10; K$\leq$5).}
   {Star catalogs available at the CDS are consulted via web
requests. They provide all the useful information for selecting of calibrators. 
Missing photometries are computed with an accuracy of 0.1~mag and the missing 
angular diameters are calculated with a precision better than 10\%. 
For each star the squared visibility is computed by taking the
wavelength and the maximum baseline of the foreseen observation into account.}
   {\emph{SearchCal} is integrated into ASPRO, the
interferometric observing preparation software developed by the
JMMC, available at the address: http://mariotti.fr. }
   {}

\titlerunning{\emph{SearchCal}: the bright object case}
\authorrunning{D. Bonneau et al.}
\maketitle

\section{Introduction}{
A long baseline optical interferometer measures the
spatio-temporal coherence (i.e. the visibility) of the target.
This measure is directly related to the Fourier transform of the
object's intensity map at the spatial frequencies
$\frac{\vec{B}}{\lambda}$, $\vec{B}$ being the baseline vector
between two telescopes. Optical interferometry provides a powerful
tool for determining the morphology of astronomical sources at high
angular resolution. The modulus and the phase of the visibility
are derived respectively from the contrast and the position of the
fringes resulting from the recombination process.

The atmospheric turbulence and instrumental instabilities induce
long-term and short-term drifts that distort the phase and
decrease the amplitude of the target's visibility $V_{target}$ by
a factor $\Gamma$ called the instrumental response. The observed
visibility $\mu_{target}$ could then be written as

\begin{equation}
    {\mu^2}_{target}={V^2}_{target} \Gamma^2.
\end{equation}

In order to take these effects into account and to convert the
observed fringe contrast into true visibility, the observation of
the scientific target is usually bracketed by observations of
calibration stars. In practice, observing a calibration star whose
visibility $V_{cal}$ can be accurately deduced from direct or
indirect determination of its angular diameter leads to a
determination of $\Gamma$:
\begin{equation}
    \Gamma=\frac{|\mu_{cal}|}{|V_{cal}|}.
\end{equation}
It is then possible to calculate the accuracy on the target's
visibility:
\begin{equation}
\frac{\Delta V_{target}^2}{V_{target}^2}\simeq \frac{\Delta V_{cal}^2}{V_{cal}^2} + \frac{\Delta\mu^2}{\Gamma^2}(\frac{1}{V_{cal}^2} + \frac{1}{V_{target}^2})
\end{equation}
where $\Delta\mu^2 \simeq \Delta{\mu^2}_{target} \simeq \Delta{\mu^2}_{cal}$ 
is the uncertainty on the measurement of the visibility amplitudes.
Equation 3 shows that the expected accuracy on the target
visibility strongly depends on the accuracy of the calibrator
visibility.

A calibrator is a star for which the visibility is known (or can
be predicted) with high accuracy. It should have physical
properties (magnitude, spectral-type, colors) and a sky location
close to those of the scientific target, so that the instrumental
response during the calibrator-target-calibrator sequence could be
considered as independent of the object.

The selection of suitable calibration stars is crucial to obtain
the ultimate precision of the interferometric instruments. Until
now, each interferometric group has had its own strategy for
calibrating the observations either by using reference stars chosen
case by case or using specific tools of selection. In
\cite{borde}, Bord\'e et al. published a catalog of $374$
reference stars selected from the initial list of Cohen's
spectro-photometric calibrators (Cohen et al. \cite{cohen}) using
selection criteria adapted to infrared interferometry up to a
$200~m$ baseline. More recently, an observing program (Percheron
et al. \cite{percheron}) was set up to create a list of
reference stars with accurate measured angular diameters suitable
for calibrating the infrared interferometric observations of the Very
Large Telescope Interferometer (VLTI) instruments (VINCI, MIDI,
AMBER). To prepare interferometric observations with the Palomar
Testbed Interferometer (PTI) and Keck Interferometer (KI), an
interferometric observation planning software \emph{GetCal} has
been developed (Boden \cite{boden}), including a tool to compute
the visibility of potential reference stars taken in the Hipparcos
catalog and extracting astronomical and spectro-photometric
parameters from the Simbad database at the Centre de Donn\'{e}es
Astronomiques de Strasbourg (CDS)(Genova et al., \cite{cds}).

With the startup of long-baseline and large-aperture optical
interferometers, such as VLTI, KI, or the CHARA array, and with the
increase in the accuracy or in the range in sensitivity and angular
resolution, the calibration of interferometric data requires
definiting new strategies for seeking suitable calibrators
and developing of selection tools usable by a larger community
of astronomers.

In Sect. 2, we present our method for creating a dynamical list
of stars fulfilling the requirements of interferometric
calibrators for a bright scientific target($K \leq 5$). Section 3 briefly
depicts the different scenarii of request to the CDS database, in
order to extract the useful parameters from stellar catalogs and
to sort out the initial list of possible calibration stars. Section 4
deals with the major steps in the calculations (interstellar
absorption, angular diameters, visibility) for each star on the
list. Some technical aspects are mentioned in Sect. 5.
Finally, the current limitations of \emph{SearchCal} and its
evolution to in case of faint targets are discussed in the last
section. }

\section{Our original method}{
The design of a search calibrator tool available in ASPRO (Duvert
et al., \cite{aspro1}, Duchene et al., \cite{aspro2}) was guided 
by the goal of creating a dynamical catalog of calibration
stars suitable for each scientific target. The goal was to provide
a list of potential calibration stars for which the visibilities
are calculated from their angular diameters and the maximum
spatial frequency ($\frac{\vec{B}}{\lambda}$) of the
interferometric observation. The search for calibrators must work
as well for long baseline interferometric observations carried out
in the visible ($V$ band), the near infrared ($J$, $H$, or $K$
bands), or the mid infrared ($N$ band).

The ``\emph{Virtual Observatory}'' techniques were adopted to
extract the required astronomical information from a set of
stellar catalogs available at the CDS. Compared to the static or
closed-list approach, the merit of this strategy is first to
take into account any enrichment of the catalogs by new
observational data and secondly to be much more adapted to the
limits in magnitude of the coming interferometric facilities (VLTI
with four $8~m$ or KI with two $10~m$ telescopes).

To minimize the effects of temporal and spatial variations of the
seeing on the calibration process, a calibrator must be as close
as possible to the scientific target. The field size on the sky is
defined by the maximum difference in right ascension and
declination. To be observable with the same instrumental
configuration, the magnitude of the calibrator must be in a narrow
range of value around the target magnitude in the observing
photometric band. In order to select stars as potential
calibrators, a certain number of astronomical parameters must be
known for each star. These parameters are given in Table~\ref{AstroParam}.

\begin{table}[h]
    \caption{Astronomical parameters for calibration stars}
    \label{AstroParam}
    \centering
        \begin{tabular}{ll}
      \hline
        Identifiers & HIP, HD, DM numbers \\
        Astrometry & coordinates (RAJ2000, DEJ2000),\\
        & proper motion, parallax,\\
        & galactic coordinates\\
        Spectral Type & temperature and luminosity class \\
        Photometry & magnitudes $U,B,V,R,I,J,H,K,L,M,N$ \\
        Angular diameter & measured or computed angular diameter \\
        Miscellaneous & variability and multiplicity flags,\\
        & radial velocity, rotational velocity \\
      \hline
        \end{tabular}
\end{table}

An on-line interface with the VizieR data base (Ochsenbein et al.,
\cite{vizier}) at the CDS was created to extract astrometric and
spectro-photometric parameters of the sources in the defined box
and to obtain the initial list of stars (see details in the next
section). This list is enriched by the stars present in the
\emph{Catalogue of calibrators for long baseline stellar
interferometry} (Bord\'e et al., \cite{borde}) and the
\emph{Catalog of bright calibrator stars for 200-m baseline
near-infrared stellar interferometry} (M\'erand et al.,
\cite{merand}). If available, the measured angular diameter is
obtained through the data of the \emph{Catalog of High Angular
Resolution Measurements} (Richichi, Percheron and Khristoforova,
\cite{charm}).

For each star on the initial list, calculations are made to
correct the interstellar absorption and to compute missing
magnitudes. The photometric angular diameter and its associated
accuracy are estimated using a surface brightness method based on the
$(B-V)$, $(V-R)$ and $(V-K)$ color index. Then, the expected
visibility and its error are computed.

The list of possible calibrators is finally proposed to the user
and the final choice can be made by changing the selection criteria:
accuracy on the calibrator visibility, size of the field,
magnitude range, spectral type and luminosity class, variability
and multiplicity flags. }

\section{The CDS interrogation}{
To build a dynamical list of stars, we chose to extract the
information from catalogs available at the CDS. Different
scenarii were implemented depending on the photometric band selected for the
interferometric observations, i.e. in the visible (V band) or the
near infrared (K band). For each star the astronomical parameters
were extracted from the following catalogs:
\begin{itemize}
\item I/280: All-sky Compiled Catalog of 2.5 million stars (Kharchenko, 2001)

\item II/7A: UBVRIJKLMNH Photoelectric Catalog (Morel et al., 1978)

\item II/225: Catalog of Infrared Observations, Edition 5 (Gezari et al., 1999)

\item II/246/out: The 2MASS all-sky survey Catalog of Point Sources (Cutri et al., 2003)

\item J/A+A/413/1037: catalog J-K DENIS photometry of bright southern stars ((S. Kimeswenger et al.,2004)

\item I/196/main: Hipparcos Input Catalog, Version 2 (Turon et al., 1993)

\item V/50: Bright Star Catalog, 5th Revised Ed. (Hoffleit et al., 1991)

\item V/36B: Supplement to the Bright Star Catalog (Hoffleit et al., 1983)

\item J/A+A/393/183: Catalog of calibrator stars for LBSI (Bord\'e et al., 2002)

\item J/A+A/433/1155: Calibrator stars for 200-m baseline interferometry (Merand et al., 2005)

\item J/A+A/431/773/charm2: Catalog of High Angular Resolution Measurements (Richichi et al., 2005)
\end{itemize}

To define of the extracted data and the limitation of the
number of returns, we used for each catalog the data fields
defined by UCDs (Unified Content Descriptors) and labels and the
limits on the data's values. Our strategy is based on two
sequences of requests on the VizieR data base. 

\subsection{Primary request}{

An on-line interface with the CDS has been created to obtain the
initial list of stars present in the calibrator field and with
magnitudes according to the specified magnitude range. This
request is done on catalog(s) called ``primary catalog(s)'' depending
on the scenario. The primary catalogs were selected with
respect to the quality of the equatorial coordinates and of the
available photometry. In the case of the ``visible'' scenario, the
choice was thus made on the compiled catalog I/280 because of
the need to have a reliable value for the magnitude $V$ and
precise coordinates for stars brighter than typically $Vmag \leq
10$. For the ``near infrared'' scenario, it was mandatory to have
the $K$ magnitude of a star brighter than typically $Kmag \leq 5$,
and the choice was made to take the compiled catalogs I/225, II/7A,
and II/246 as primary catalogs. The output of this first sequence
of requests is a list of star coordinates having magnitude values
as specified in the defined calibrator field. }

\subsection{Secondary request}{
The second sequence of requests is done on the stars contained in
the previous list and with the goal of extracting astrometric and
spectro-photometric parameters. The secondary catalogs were
selected because of the relevance and the reliability of the
parameters of interest for our purpose. In the current version of
\emph{SearchCal}, the identifiers, the equatorial coordinates, the
proper motions, the parallaxes, spectral type, and variability
or multiplicity flags are extracted from the I/280 catalog. For
bright stars, the visible photometry comes from I/280, whereas
infrared photometry is taken from the II/7A, II/225 and II/246
catalogs. The galactic coordinates are taken from the I/196 or II/246
catalogs. The radial velocity and rotational velocity are
extracted from the I/196, V/50 or V26B catalogs, respectively. }

\subsection{Setting the list of possible calibrators}{
We then parse and merge all the results in a single array of stars
with all the astronomical parameters. The catalogs are first linked
according to the HD number if provided and with the
equatorial coordinates found in the different catalogs if they are
coherent at the level of 1 arc second. The $V$ magnitude (for
$V$ band) or the $K$ magnitude (for infrared bands) is also used
to confirm that the star present in the different catalogs is the
same.
The final result is a single list containing stars for which the suitable
astronomical parameters have been extracted from the selected catalogs. }
}

\section{The central engine of the calibrator's parameters calculation}
For the stars contained in the final list of the CDS requests, we
need to compute their apparent diameters (except if they have already been
measured) to determine their visibilities in the interferometric
configuration. This is done in several stages: first we correct
the photometry from interstellar absorption, then we compute the
possible missing photometric data, and finally the apparent
diameter is obtained from surface brightness relation.

\subsection{Interstellar absorption}{
We must correct the photometric data for the wavelength-dependent
effects of the galactic interstellar extinction. In the current
version of \emph{SearchCal}, all the calibrators are bright enough
to have a measurement of their trigonometric parallax. For each
star, the visual absorption $A_V$ is computed as a function of the
galactic coordinates (longitude $l$ and latitude $b$) and of the
distance $d$. As all calibration stars currently selected by
\emph{SearchCal} have $d \leq 1000 pc$, we have used the analytic
expression for the interstellar extinction in the solar
neighborhood given by Chen et al. (\cite{chen}).

The observed magnitudes $mag[\lambda]$ are then corrected for
interstellar absorption using:
\begin{equation}
$$mag[\lambda]_0 = mag[\lambda]- A_{\lambda}$$
\end{equation}

\begin{equation}
$$R_{\lambda} = A_{\lambda}/E(B-V)$$
\end{equation}

\begin{equation}
$$A_{\lambda} = A_V R_{\lambda}/R_V$$
\end{equation}
with $R_V = 3.10$ and the values of $R_{\lambda}$ given by
Fitzpatrick (\cite{fitzpatrick}). }

\subsection{Rebuilding missing photometries}{
The knowledge of all the $BVRIJHK$ photometry of the calibrator is useful 
for the interferometric observation. 
In particular, it could be important to have a calibrator with similar brightness 
to the scientific target at the observing wavelength, and the photometry will be  
mandatory for computing the angular diameter of the calibrator. 
In addition, the value of the magnitude of the calibrators in some photometric bands 
must be known for the needs of some housekeeping operations (fringe tracking for example).

If some photometric values are missing, we have chosen to compute them from
``spectral type - luminosity class - color'' relations and from
existing magnitudes. Such relations exist in the literature, but
in general each of them covers only certain classes of luminosity,
a range of spectral type, or a limited number of photometric
index. To obtain a relation linking all spectral types of all
luminosity classes to $BVRIJHKLM$ photometry, we compiled the
works of Bessel (\cite{bessel79}), Bessel and Brett
(\cite{bessel88}), Fitzgerald (\cite{fitzg70}) Johnson
(\cite{johnson}), Leggett (\cite{legett}) Schmidt-Kapler et al.,
(\cite{schmidt}), Th\'e (\cite{the}) and Wegner (\cite{wegner}).
We adopted the Johnson photometric system according to our
main source of accurate photometry (Morel \& Magnenat
\cite{morel}). The relation of Bessel (\cite{bessel83}), Glass
(\cite{glass75}) and Bessel and Brett (\cite{bessel88}) are used
to transform the other photometric systems in the Johnson one. 
Tables~\ref{tab_phot_dwarfs}, \ref{tab_phot_giants}, and
\ref{tab_phot_supergiants} list the adopted value of our ``spectral
type - luminosity class - color'' relations for
dwarfs, giants and supergiants stars, respectively.
In Fig. \ref{color_spectral} we show an example of our
relation for the $(B-V)$ and $(V-R)$ color index. The $(B-V)$
relation is also used to check the consistency of the spectral
type extracted at the CDS.

\begin{figure}
\begin{tabular}{c}
\includegraphics[width=6.3cm,angle=-90]{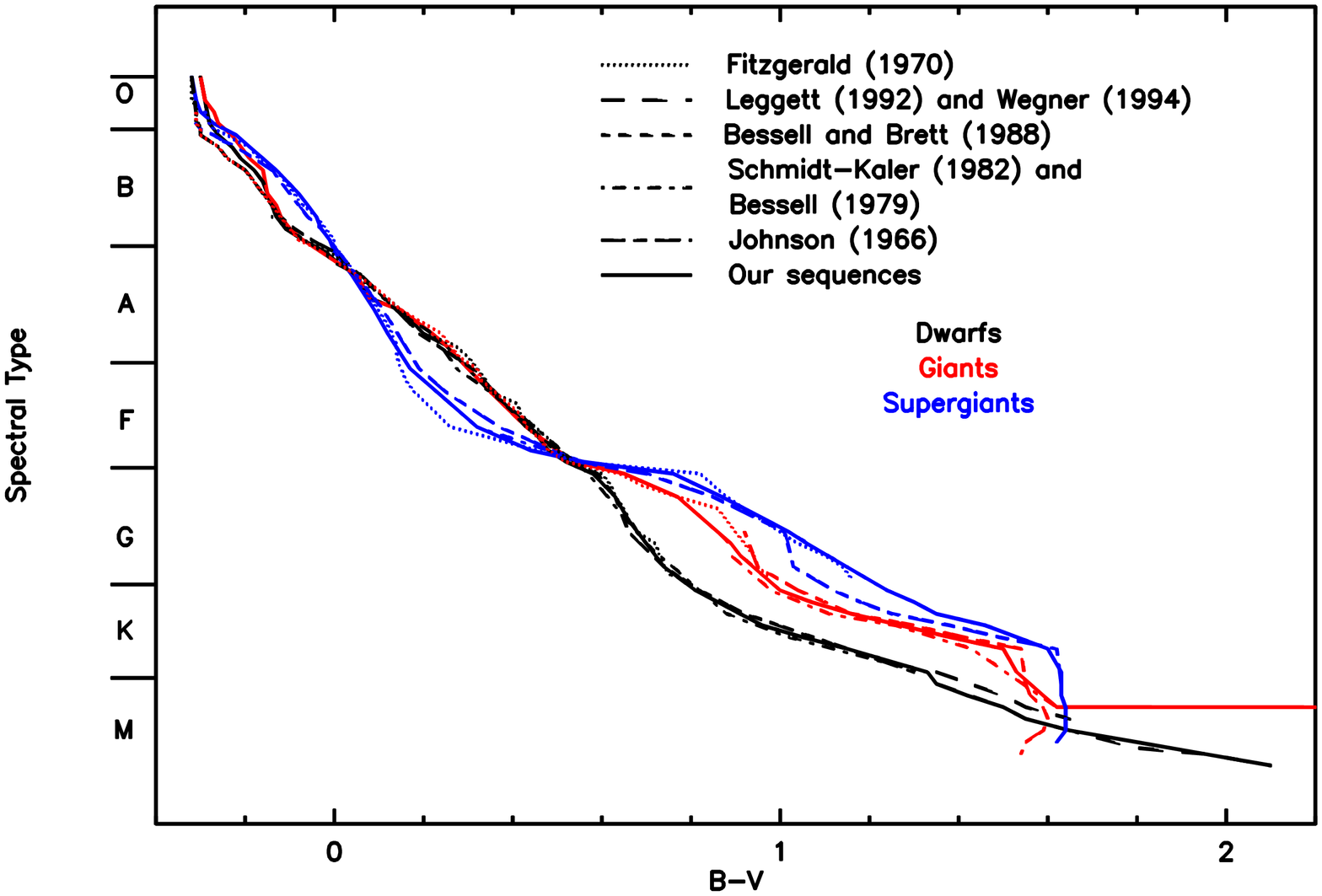} \\
\includegraphics[width=6.3cm,angle=-90]{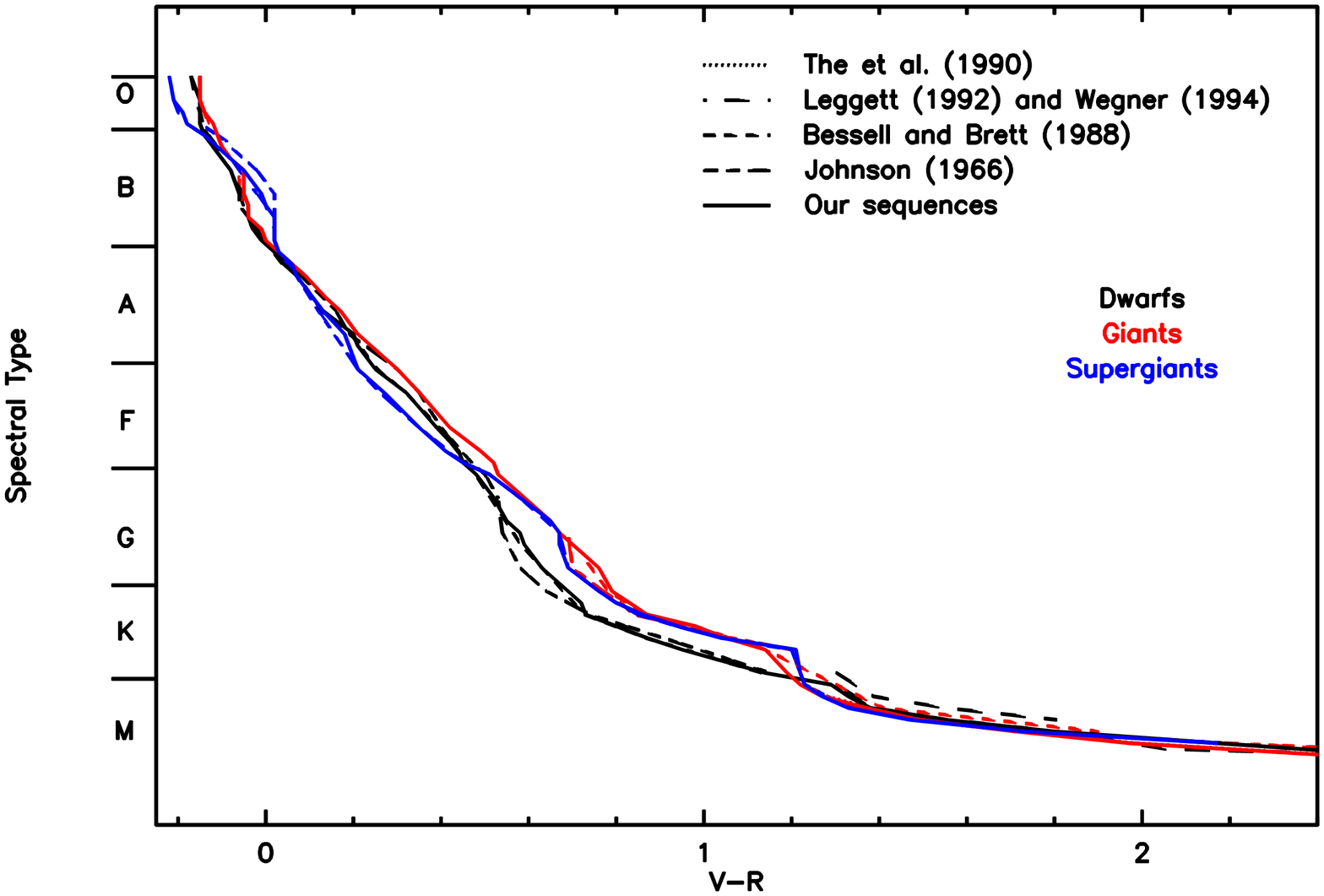}
\end{tabular}
\caption{Example of our ``spectral type - luminosity class - color''
relations used to compute missed photometry. The relation for
dwarfs, giants and supergiants are shown in dark,
red, and blue, respectively . Our sequences are plotted in solid lines, when the
relation from which they are extracted are in different dashed
lines.} \label{color_spectral}
\end{figure}

To check the accuracy of the rebuilt photometry, we compare the
colors of our tables with those of stars in the catalog of
Ducati (\cite{ducati02}) also in Johnson filters. 
Figure~\ref{accuracy_color} shows an example of the O-C
(difference between the color measured and computed with our
relation) as a function of the spectral type. The Ducati
(\cite{ducati02}) photometry is not corrected for
interstellar absorption, so a part of the dispersion is due to
reddening, which is visible for the bright (and then distant) OB
stars. The dispersion of the O-C is then a superior limit of
the accuracy of our computed photometry. Using only the stars in
the spectral type range A to K seems a good compromise for estimating
our accuracy, since they are closer than OB stars so less
reddened and enough bright to reduce the observational
errors.

In the case of near infrared observations, we impose the knowledge
of the photometry in $V$ and $K$, and our computed complementary
photometry has an accuracy of 0.1~mag or better. For visible
observations, only the $B$ and $V$ magnitudes are mandatory, so that
the determination of the missing $J$, $H$, and $K$ has an accuracy
of 0.2~mag.

\begin{figure}
\begin{tabular}{c}
\includegraphics[width=6.3cm,angle=-90]{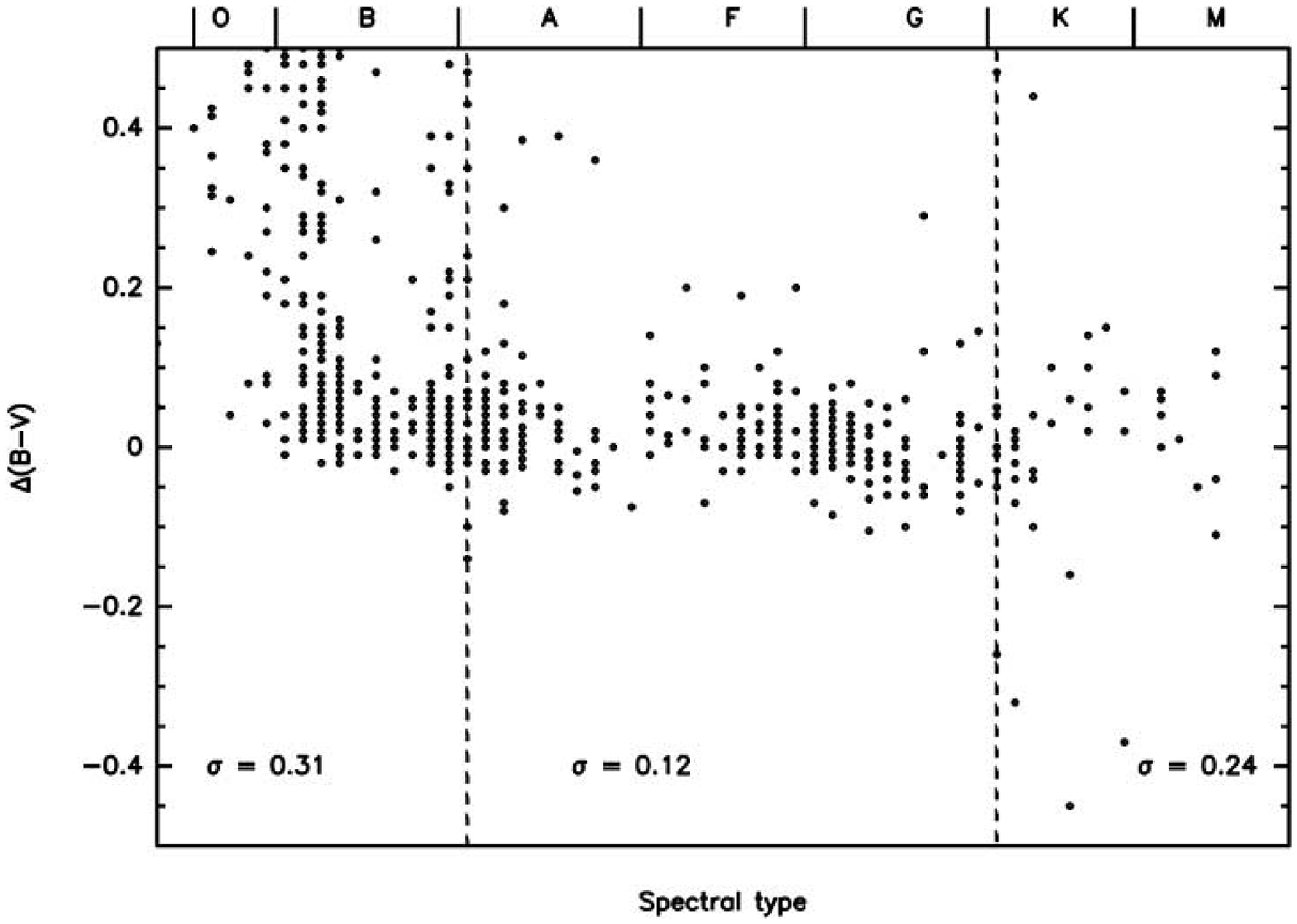} \\
\includegraphics[width=6.3cm,angle=-90]{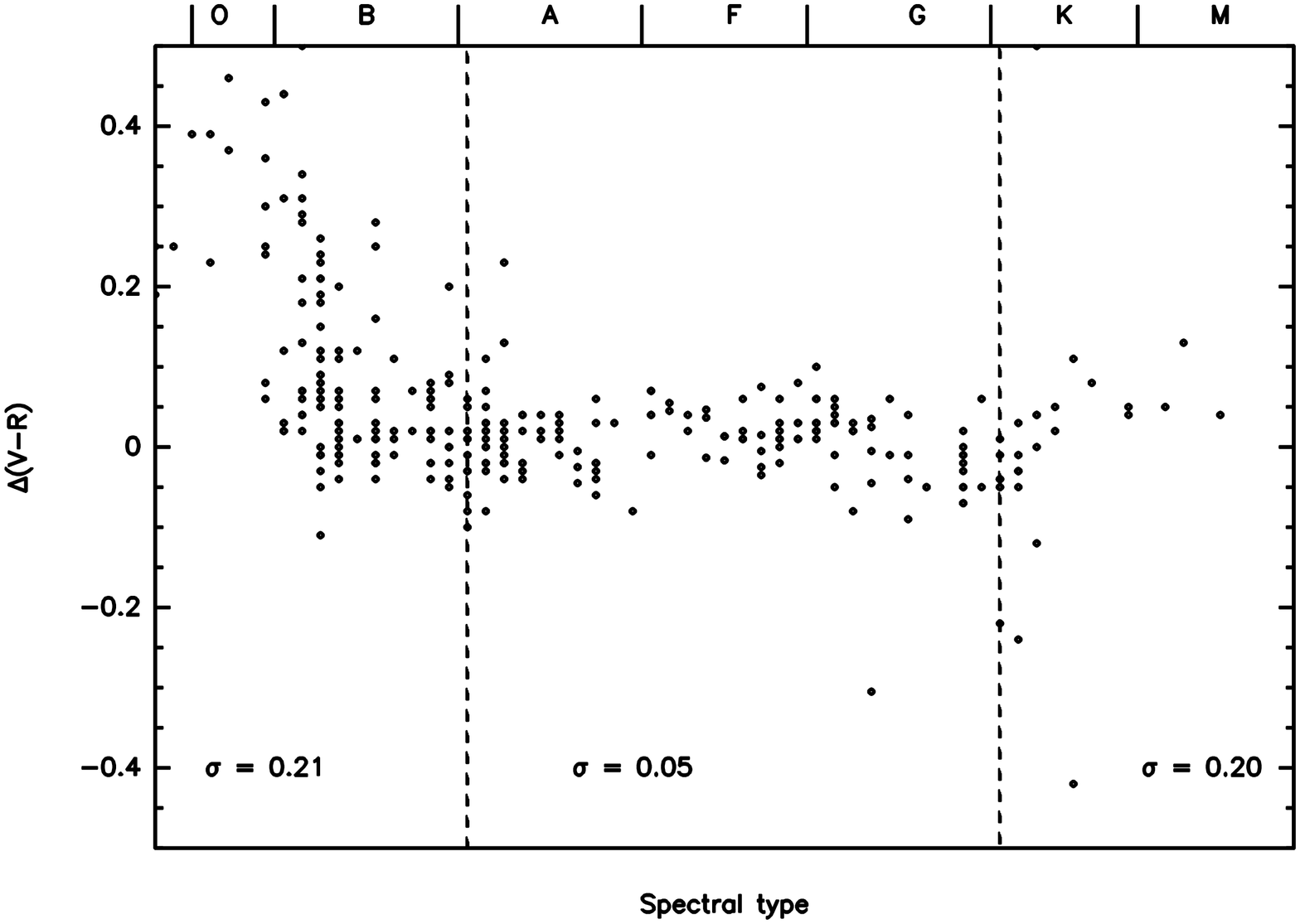}
\end{tabular}
\caption{
Difference between measured colors of the stars in the Ducati catalog
 (\cite{ducati02}) and that computed for the same
spectral type and luminosity class. The rms of this O-C
are given for three ranges of spectral type. Only the dwarfs are plotted
in this figure, as an example.}
\label{accuracy_color}
\end{figure}

\subsection{Determination of the angular diameter}
Since the goal was to calculate the visibility for each
possible calibrator, it was necessary to know the value of its
angular diameter. In some particular cases this parameter is
either a measured value taken from catalog CHARM (Richichi et
al., \cite{charm}) or an estimated value taken from the lists of
stars of reference published by Bord\'e et al. (\cite {borde}) or
M\'erand et al. (\cite {merand}). These catalogs are indeed
included in {\it SearchCal}. But in the general case no object in
these catalogs complies with the specific requests of ASPRO in
coordinates and magnitude range, and then the angular diameter is
usually not known. A surface-brightness versus color-index
relation should be used to compute angular diameter from
photometry. Such relations exist in the
literature (di Benedetto \cite{bene98}; van Belle \cite{belle99};
Kervella et al. \cite{ker04}) but are determined only for
a particular luminosity class or only for few a photometric indices.

Our goal was then to obtain a universal relation working for the all 
luminosity classes. 
For that purpose we used, on one hand, linear diameters and an absolute magnitude 
determined in eclipsing binaries (which are in general dwarfs) and, 
on the other, angular diameters measured from interferometry or 
lunar occultation (generally for giants) and apparent magnitude. 

The angular diameter $\theta$ of a star of linear diameter $D_*$ 
(in unit of solar diameter $D_\odot$) at the distance $d$ (in pc) 
is given by:

\begin{equation}
\theta = Cst \frac{D_*}{d} 
\end {equation}
where Cst = 9.306 mas corresponds to the angular diameter of the sun seen at 1 pc.
The distance of the star is usually a function of the apparent $m_V$ 
and absolute $M_V$ magnitudes:

\begin {equation}
d = 10^{(m_V - M_V +5)/5}.
\end {equation}

Then, we define the quantity $\psi_V$ as:

\begin{equation}
\psi_V = \frac{D_*}{10^{(5-M_V)/5}} =
\frac{\theta}{9.306.10^{-m_V/5}},
\label{esti_dia}
\end{equation}
where $\psi_V$ is computed as a function of $\theta$ and $m_V$ for
stars with angular diameter measured from interferometry or 
lunar occultation and as function of $D_*$ and $M_v$ for 
eclipsing binary components.
Then, $\psi_V$-versus-color-index relations were determined for 
the whole index of $BVRIJHK$ system using a polynomial fit 
for each color index ($CI$).

\begin{equation}
\psi_V =  \Sigma_k a_k CI^k.
\end{equation}

To determine our relations we compiled (i) the 
stellar diameter (from interferometric measurements, lunar
occultation and eclipsing binaries) from Barnes et al.
(\cite{barne78}), Andersen (\cite{andersen91}), S\'egransan et al.
(\cite{seg03}), and Mozurkewich et al. (\cite{mozu2003}) and (ii) 
$BVRIJHK$ photometry (from Ducati \cite{ducati02} and Gezari et
al. \cite{gezari99} catalogs) for a large sample of stars of
spectral type O to M and for the whole luminosity class. We plan to
regularly add new and more accurate measurements in our compilation, 
and to refresh our relation in {\it SearchCal}. Such
relations are also very useful for other studies and will be
described in detail in a forthcoming paper.

\begin{figure}
\begin{tabular}{c}
\includegraphics[width=6.3cm,angle=-90]{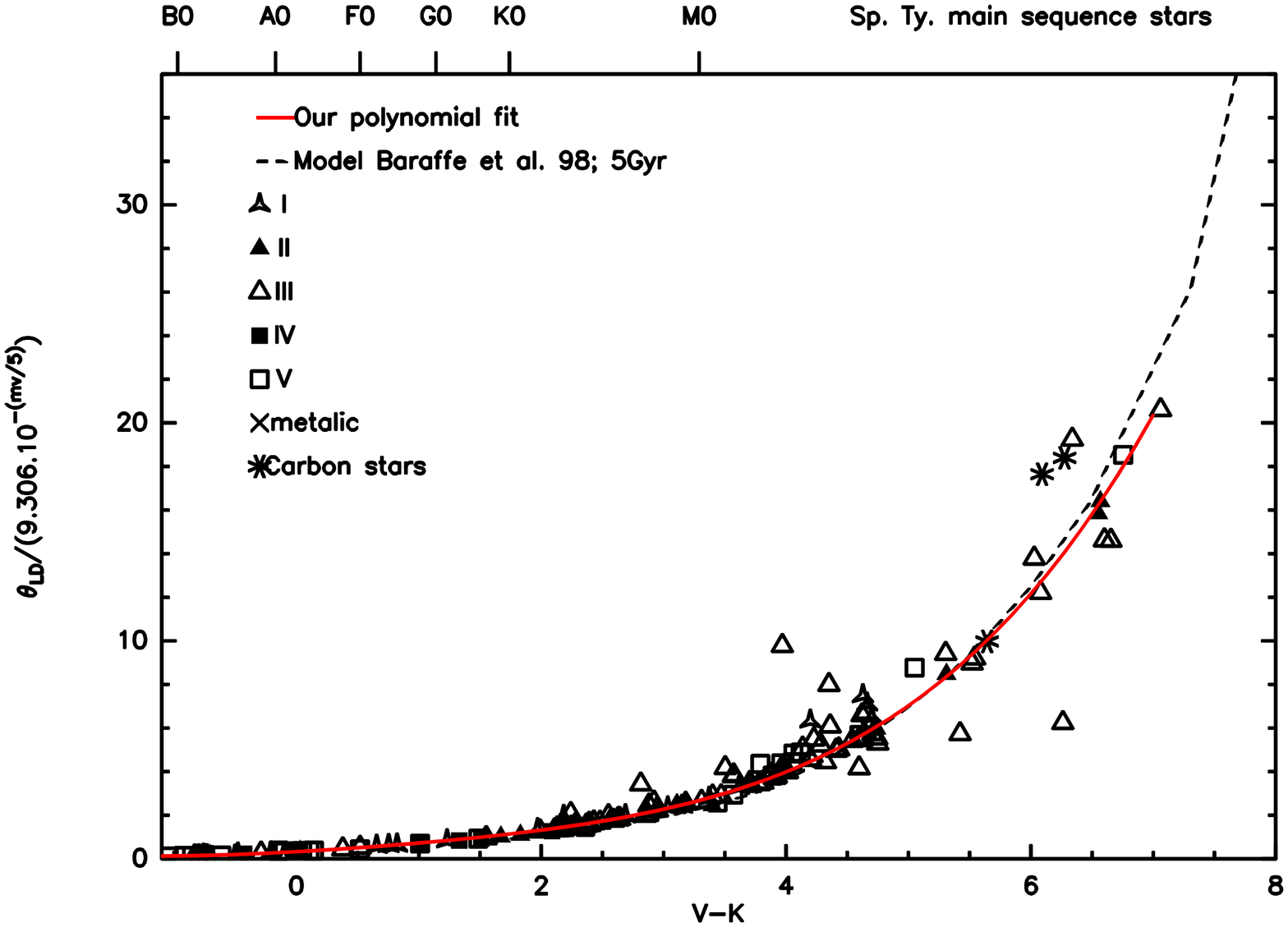} \\
\includegraphics[width=6.3cm,angle=-90]{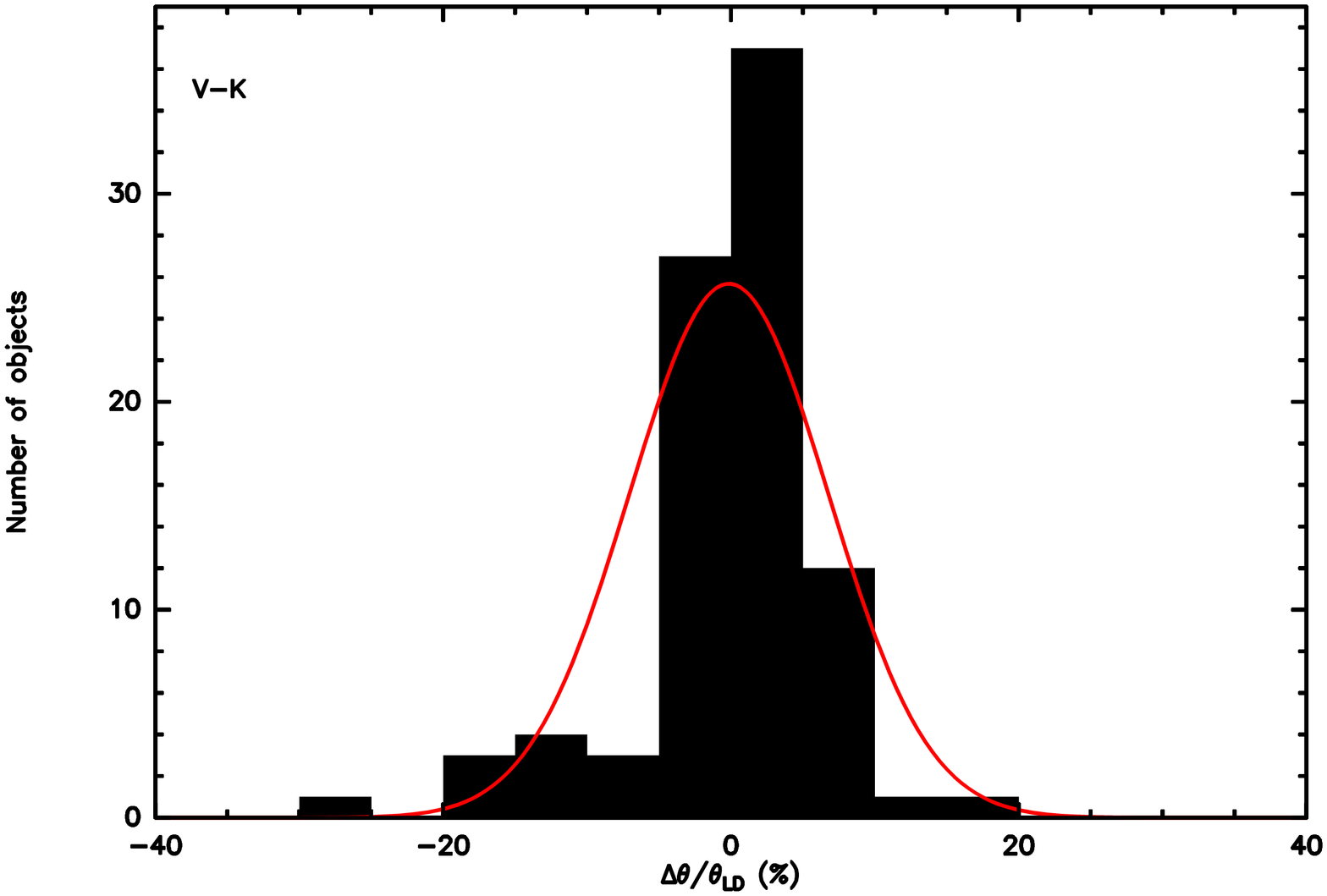}
\end{tabular}
\caption{ top: $\psi_V$ as a function of $(V-K)$ color.
The red line is our polynomial fit.
bottom: Distribution of the angular diameters (O-C) from the
$\psi_V$ versus $(V-K)$ colors relation. The red curve
is a Gaussian function fitting the distribution of the (O-C).
$\Delta \theta$ is the difference between angular diameters
computed from our relation and measured angular diameters} 
\label{fig_psi}
\end{figure}

In the top of Fig.~\ref{fig_psi} we plot one example of the relation for the
$(V-K)$ color. The angular diameters could then be
computed by using Eqs.~9 and ~10. The (O-C) (difference between
the measured and the computed angular diameters) are calculated
for the stars of S\'egransan et al. (\cite{seg03}) and Mozurkewich
et al. (\cite{mozu2003}). The distribution of the relative (O-C)
are shown in the bottom of the Fig.~\ref{fig_psi}.

The first cause of uncertainty in the computation of the angular
diameter is the variance of the calibration residuals (variance of
the (O-C)). This variance includes the intrinsic dispersion of the
relations, the error on the measured diameters and the error on the
photometry. The three relations with the best accuracy are
$\psi_V(B-V)$, $\psi_V(V-R)$, and $\psi_V(V-K)$ with
an uncertainty of 8\%, 10\%, and 7\%, respectively. We choose these
three relations to determine the stellar angular diameter in {\it
SearchCal}. The polynomial fit of these three relations are given
in Table~\ref{poly_psi}.

\begin{table*}
\begin{center}
\begin{tabular}{lllllllll} \hline
 col. Ind. &  Validity domain &  a$_0$ &  a$_1$ &  a$_2$ &  a$_3$ &  a$_4$ &  a$_5$ & Accuracy \\ \hline
 B-V &  [-0.4; 1.3]  &  0.33822617  &  0.76172888  &  0.16990933  &  -0.0803159  &  0.36842746  &  ----       & 8\%   \\
 V-R &  [-0.25; 2.8] &  0.29974514  &  0.90469909  &  -0.0438167  &  2.32526422  &  -1.4324917  &  0.43618476 & 10\%  \\
 V-K &  [-1.1; 7.0]  &  0.32561925  &  0.31467316  &  0.09401181  &  -0.0187446  &  0.00818989  &  ----       & 7\%   \\ \hline
\end{tabular}
\end{center}
\caption{Polynomial coefficient of the $\psi_V =  \Sigma_k a_k
CI^k$ relation for the three color-index retained. This relation
is only defined for a given validity domain in color.}
\label{poly_psi}
\end{table*}

The error in the photometry is propagated in the angular diameter
when a surface brightness relation is used. As already mentioned,
we impose the restriction that the $K$ magnitude of the calibrators are from
observations and that the stars are in the Kharchenko (2001)
catalog, where the $B$ and $V$ magnitudes are present. Thus, for the
three colors used to determine the angular diameter, only the
$(V-R)$ color is determined from our spectral type - color
relation, then could have substantial error. In Fig.~\ref{err_vr} 
we show the propagation of the errors in
$(V-R)$ on the angular diameter.

\begin{figure}
\includegraphics[width=6.3cm,angle=-90]{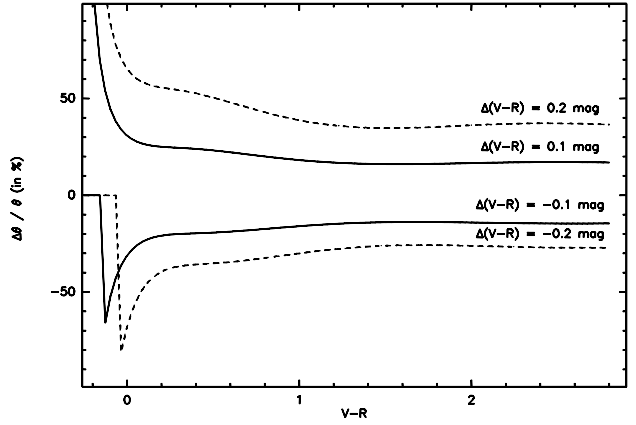}
\caption{Error in the determination of the angular diameter when
the $\psi_V$ versus $(V-R)$ relation is used and when the $(V-R)$
color have errors of -0.2, -0.1, 0.1, and 0.2 magnitude.}
\label{err_vr}
\end{figure}
In conclusion, our angular diameter determination has an accuracy
of $\leq$10\% if measured photometric data are used.
When the angular diameter is computed with a
$(V-R)$ calculated data, the accuracy is $\sim$20\%.

For each star, a coherence test of the photometry is done by
comparing the computed diameters with the different color
indexes, $\theta[BV]$, $\theta[VR]$, $\theta[VK]$. The star is
rejected from the list if one value of the angular diameter
differs from more than $2\sigma$ from the mean value.

In the top of Fig. 3, the enhanced scattering of stars with $(V-K) > 4$ 
(spectral types later than M0) around the polynomial fit mainly reflect 
an intrinsic dispersion of the photometric data for cool evolved stars. 
This can be related to stellar variability or to the presence of a 
circumstellar envelope and a color dependent variation 
of the computed angular diameter results. Then this type of star cannot be 
considered as a good calibrator, so it is rejected from the final calibrator list.

\subsection{Computation of the squared visibility}

The visibility of the calibrator must be computed rigorously to avoid 
any differential effect with the visibility of  the scientific target.
For wide spectral band interferometry, a monochromatic visibility can be computed 
for an  effective wavelength taking the spectral-energy distribution of the star into account (calibrator or scientific target) across the filtered spectral bandwidth. 
For spectrally resolved interferometric measurements (wavelength-resolved fringes), 
the polychromatic visibility must be computed across the observed spectral bandwidth.
In this version of SearchCal, we give only an estimation of the visibility for 
the central wavelength of the photometric band used for the observation.  
Each star in the list is considered as a uniform disc. The
squared visibility $V_{cal} ^2$ and its associated error $\Delta
V_{cal} ^2$ are computed as a function of the angular diameter
$\theta$(mas) and its error $\Delta \theta$ for the given
instrumental configuration (wavelength $\lambda$(nm) and maximum
baseline $B_{max}$(m)):

\begin{equation}
        $$V_{cal} ^2$$ = $$|$$2 J_1(x)/x$$|$$^2$$
\end{equation}

\begin{equation}
        $$\Delta V_{cal} ^2$$= $$8 J_2(x)|$$J_1(x)/x$$|$$\Delta \theta/\theta$$
\end{equation}
with $x = 15.23 B_{max} \theta/\lambda$.

To compute the visibility, the value of the diameter can either be the measured $\phi_{ud}$ 
or $\phi_{ld}$ taken from the catalog CHARM, the computed $\phi_{ld}$  
given in the Catalog of Calibrators for Long Baseline Stellar Interferometry, 
or the photometric angular diameter $\phi[VK]$.
Using $\theta_{ld}$ instead of $\theta_{ud}$ in Eq.[11] induces a bias 
in the computed visibility,
$\delta V^2 = {V_{cal}} ^2(\theta_{ud}) - {V_{cal}}
^2(\theta_{ld})$ which must be estimated.
With the linear representation of the limb darkening, a good
approximation of the ratio $\theta_{ld}$/$\theta_{ud}$ as function of
the limb darkening coefficient $u$ is given
by:
\begin{equation}
    $$\theta_{ld}$/$\theta_{ud}$$=$$[$$(1-u/3)/(1-7u/15)$$]^2$$
\end{equation}
with $0.0 \leq u \leq 1.0$ and then $1.0 \leq
\theta_{ld}/\theta_{ud} \leq 1.12$.
The bias on the visibility is maximum for a full darkened disk ($u
= 1.0$) and it is always less then $5\%$ for $x < 1.0$ and
$V_{cal}^2 > 0.75$ (see Table~\ref{BiasVisi}).

\begin{table}[h]
    \caption{$x_{max}$ as a function of the maximum bias $\delta V^2$ on the computed visibility}
    \label{BiasVisi}
    \centering
        \begin{tabular}{ll}
            \hline
        $\delta V^2_{max}$&$x_{max}$ \\
      \hline
        0.05 & 1.0 \\
        0.025 & 0.65 \\
        0.01 & 0.4 \\
      \hline
        \end{tabular}
\end{table}
Then, for a requested accuracy $\Delta V^2$ of the computed
visibility, the bias $\delta V^2$ can be neglected if the angular diameter 
of the potential calibrator satisfied the condition:

\begin{equation}
    \theta(mas) \leq x_{max}(\delta V^2)\lambda(nm)/15.23 B(m)
\end{equation}
Figure \ref{diamax} shows the values of $\theta(mas)$ as function of the base length $B$
and of the maximum bias $\delta V^2$ for the $K$  band ($\lambda = 2.2 \mu m
$).

\begin{figure}
\resizebox{\hsize}{!}{\includegraphics{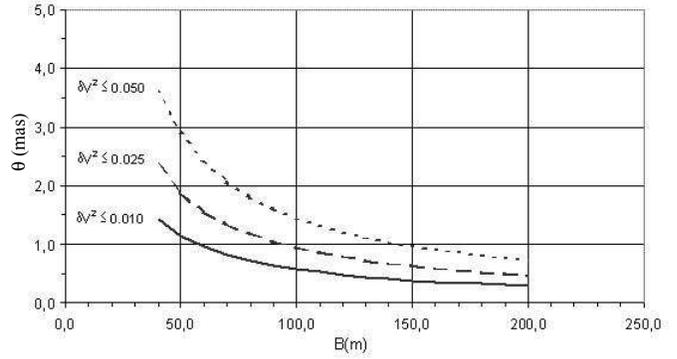}}
\caption{The maximum value of the angular diameter allowed for a
calibrator as a function of the baseline for different values of the
expected visibility bias at $\lambda 2.2 \mu m $.} \label{diamax}
\end{figure}
}

\section{Technical aspects}
\emph{SearchCal} has been designed as a distributed application that:
\begin{itemize}
\item retrieves user requests (either through a GUI interface or via an command line);

\item shifts through CDS-based stellar catalogs to retrieve a large number
of stellar parameters, according to various scenarii;

\item computes the missing photometry, if necessary using the
relations mentioned in Sect. 4;

\item classifies stars as potential calibrators;

\item presents the list to the user through a GUI interface and
handles further requests for search refinement and sorting.
\end{itemize}

\emph{SearchCal} reads and writes star catalog data in VO-Table format
(standardized XML-based format defined for the exchange of tabular data
in the context of the Virtual Observatory and return format of the CDS requests).
The resulting VO-Tables are parsed to select the possible calibration stars
using the ``libgdome'' XML parser.
Its Graphical User Interface (GUI) is a Java applet running on the
client side in any web browser. It has been developed using XML to
Java Toolkit and is fully integrated in the JMMC's ASPRO Web
software, allowing the user to display, sort, filter and save the
catalog of calibration stars. The server side application is
written in C++ using a flexible and scalable object-oriented
methodology. The design allows easily the application to be updated to
follow the improvements in scientific knowledge (in, e.g., the
scenarii), as well as changes in the web queries and evolutions in data formats.

The input panel of \emph{searchCal} is presented in Fig.~7. As
already described, the parameters of the request are: the
observing wavelength, the range of magnitude of calibrators in the
observing photometric band, and the field (maximum distance of
calibrators in right ascension and declination). The output panel
is presented in Fig.~8. In the ``result'' window, a summary of the
output of the request is given as three numbers: the number of
stars returned from CDS request as potential calibrators, the
initial number of potential calibrators selected after the
coherence test of the photometry, and finally the proposed
calibrators without variability or multiplicity indications in the
Hipparcos catalog. The final list of calibrators is displayed in
the central window. The origin of each parameter is encoded by a
color code corresponding to the name of the relevant catalog. For
the computed parameters (missing magnitude, angular diameter,
visibility), the color code indicates the confidence
level based on the quality of the photometric data. The detailed
description of the tables, as well as the function of the different
buttons, are given in the \emph{SearchCal} Help available as a PDF
file in the ASPRO web site.

Finally, from the final list of potential calibrators, one can
refine the selection of the calibrators by changing, \emph{a
posteriori}, the parameters of the request: field around the
scientific target, object - calibrator magnitude difference,
spectral type and luminosity class, accuracy on the calibrator
visibility, indication of variability, or multiplicity.

\begin{figure}
\resizebox{\hsize}{!}{\includegraphics{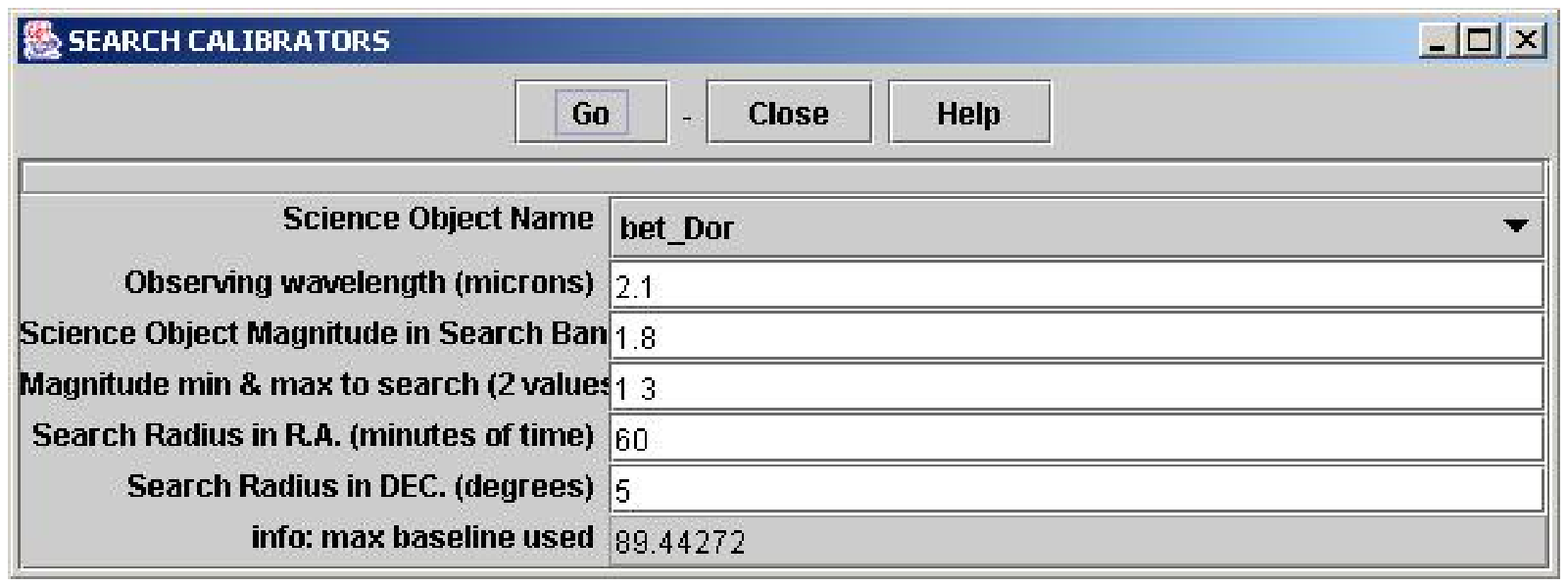}}
\caption{Input panel of SearchCal}
\end{figure}

\begin{figure}
\resizebox{\hsize}{!}{\includegraphics{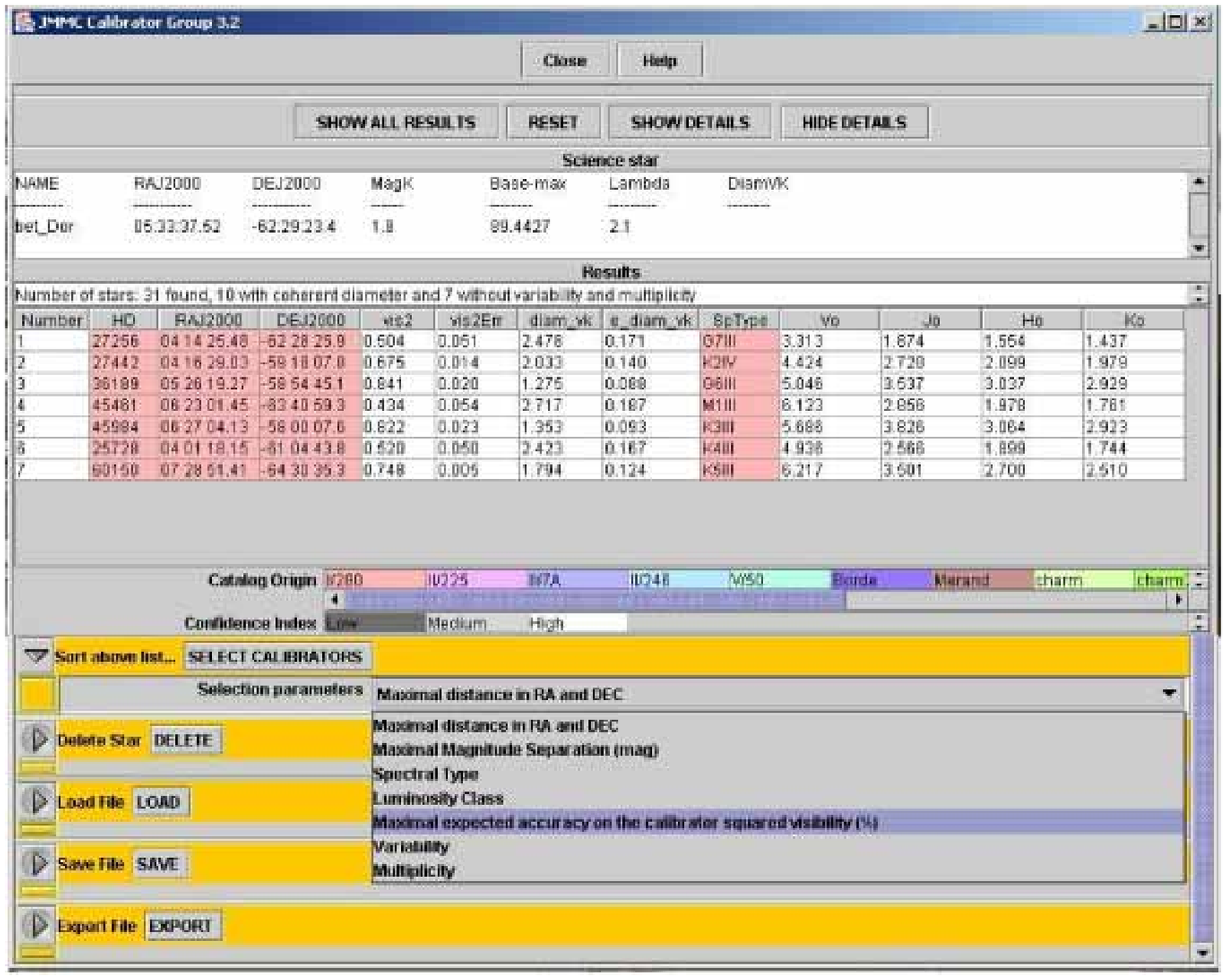}}
\caption{Output panel of SearchCal}
\end{figure}

\section{Conclusion} {
We have described the principles of our calibrator's selection
tool dedicated for optical interferometric observations. Based on
an online CDS request and a dedicated computing program, we
built a powerful piece of software that is already open to the astronomical
community. Our application (\emph{SearchCal}) and the concurrent
one (\emph{GetCalWeb}) are the only software able to find suitable
calibration stars in the vicinity of bright scientific targets and 
that are based on a dynamical approach using a Web-based
interface. However, differences of strategy and method can be noted
between these two softwares. In the current version of
\emph{SearchCal}, we impose the knowledge of the magnitudes V and
K, whereas in \emph{GetCalWeb}, the magnitude K is deduced from 
magnitude V and the spectral type. The angular diameters are
calculated in \emph{SearchCal} by a surface brightness method
whereas in \emph{GetCalWeb}, they are calculated as black bodies of
the selected spectral type and of magnitude V.

In \emph{SearchCal}, the limiting magnitude attainable for the
selected calibrators is imposed by the magnitude of the fainter
stars for which the maximum number of astrometric and spectro-photometric
parameters are available in the catalogs used for this selection,
i.e. typically $V$ magnitude $\leq 10$ or $K$ magnitude $\leq 5$.
In practice, these limits agree with the sensitivity of the
interferometers currently in operation in the visible or the near
infrared. With the gain in sensitivity expected with the
instruments AMBER and PRIMA on the VLTI, it will be necessary to
find fainter calibrators. We have already started the development of an
extended version of \emph{SearchCal} for $K$ magnitude $>5$.

The use of UCD and VOtable, as well as the splitting of our API in
three modules (``Access to CDS'', ``Computation'', ``Display''), will
allow us to easily continue the development in the framework of the Virtual
Observatory concept. Development of a CDS web service and display
in the environment of a VO portal are foreseen in near future.}

\begin{acknowledgements}{
This research has made use of the Simbad database, operated at the
Centre de Donn\'ees Astronomiques de Strasbourg (CDS), France.
This work was supported and funded by the GDR 2596
``Centre Jean-Marie Marriotti'' (JMMC) of the CNRS/SDU.
}
\end{acknowledgements}

\appendix
\section{Spectral type - luminosity class - color relations}{
\begin{table}
\tabcolsep 1.89mm
\begin{tabular}{|r|rrrrrrrr|} \hline
Sp Ty   &B-V    &V-I    &V-R    &I-J    &J-H    &J-K    &K-L    &L-M \\
\hline \hline
O5   &  -.30  &   -.42  &   -.17  &   -.31   &  -.08  &   -.14  &    .01  &   -.10 \\
O7   &  -.29  &   -.42  &   -.15  &   -.30   &  -.08  &   -.13  &    .02  &   -.09 \\
O8   &  -.28  &   -.41  &   -.15  &   -.28   &  -.09  &   -.14  &    .03  &   -.07 \\
O9   &  -.28  &   -.39  &   -.15  &   -.22   &  -.11  &   -.17  &    .03  &   -.09 \\
B0   &  -.26  &   -.37  &   -.14  &   -.21   &  -.11  &   -.17  &    .02  &   -.08 \\
B1   &  -.23  &   -.33  &   -.12  &   -.18   &  -.09  &   -.13  &   -.01  &   -.05 \\
B2   &  -.21  &   -.29  &   -.10  &   -.19   &  -.03  &   -.10  &    .00  &   -.05 \\
B3   &  -.18  &   -.24  &   -.08  &   -.15   &  -.05  &   -.09  &    .00  &   -.05 \\
B4   &  -.16  &   -.20  &   -.07  &   -.14   &  -.04  &   -.09  &    .01  &   -.04 \\
B5   &  -.15  &   -.19  &   -.05  &   -.15   &  -.05  &   -.08  &    .01  &   -.02 \\
B6   &  -.14  &   -.18  &   -.04  &   -.13   &  -.03  &   -.07  &    .03  &   -.03 \\
B7   &  -.13  &   -.16  &   -.04  &   -.12   &  -.03  &   -.06  &    .03  &   -.02 \\
B8   &  -.11  &   -.11  &   -.03  &   -.12   &  -.01  &   -.03  &    .03  &   -.02 \\
B9   &  -.07  &   -.08  &   -.01  &   -.07   &   .00  &   -.01  &    .03  &   -.01 \\
A0   &  -.02  &   -.01  &    .02  &   -.06   &  -.01  &   -.01  &    .03  &    .00 \\
A1   &   .01  &    .03  &    .04  &   -.02   &   .00  &    .00  &    .03  &    .00 \\
A2   &   .05  &    .07  &    .07  &    .02   &   .01  &    .01  &    .04  &    .01 \\
A4   &   .08  &    .24  &    .11  &    .03   &   .04  &    .05  &    .05  &    .02 \\
A5   &   .15  &    .33  &    .13  &    .03   &   .06  &    .08  &    .05  &    .03 \\
A7   &   .20  &    .30  &    .20  &    .07   &   .08  &    .10  &    .06  &    .03 \\
A8   &   .25  &    .34  &    .21  &    .09   &   .11  &    .12  &    .06  &    .03 \\
F0   &   .30  &    .42  &    .25  &    .13   &   .12  &    .15  &    .06  &    .03 \\
F2   &   .35  &    .51  &    .32  &    .14   &   .15  &    .19  &    .06  &    .03 \\
F5   &   .44  &    .68  &    .39  &    .14   &   .22  &    .26  &    .07  &    .02 \\
F7   &   .48  &    .79  &    .44  &    .16   &   .27  &    .33  &    .07  &    .02 \\
F8   &   .52  &    .81  &    .45  &    .19   &   .28  &    .34  &    .07  &    .02 \\
G0   &   .58  &    .84  &    .48  &    .21   &   .29  &    .35  &    .08  &    .01 \\
G2   &   .63  &    .87  &    .52  &    .24   &   .31  &    .36  &    .08  &    .01 \\
G4   &   .66  &    .91  &    .55  &    .24   &   .32  &    .38  &    .08  &    .01 \\
G5   &   .68  &    .94  &    .58  &    .23   &   .34  &    .40  &    .08  &    .00 \\
G6   &   .70  &    .97  &    .59  &    .24   &   .36  &    .43  &    .08  &    .00 \\
G8   &   .74  &   1.06  &    .63  &    .25   &   .39  &    .48  &    .09  &    .00 \\
K0   &   .81  &   1.14  &    .69  &    .30   &   .44  &    .53  &    .09  &   -.01 \\
K1   &   .86  &   1.20  &    .72  &    .31   &   .46  &    .57  &    .10  &   -.01 \\
K2   &   .91  &   1.26  &    .73  &    .34   &   .49  &    .59  &    .10  &   -.02 \\
K3   &   .96  &   1.38  &    .80  &    .37   &   .53  &    .62  &    .11  &   -.03 \\
K4   &  1.05  &   1.48  &    .87  &    .40   &   .57  &    .67  &    .12  &   -.04 \\
K5   &  1.15  &   1.58  &    .95  &    .43   &   .60  &    .71  &    .13  &   -.01 \\
K7   &  1.33  &   1.86  &   1.14  &    .43   &   .64  &    .78  &    .14  &    .06 \\
M0   &  1.35  &   2.10  &   1.29  &    .45   &   .73  &    .92  &    .21  &    .10 \\
M1   &  1.42  &   2.31  &   1.33  &    .52   &   .72  &    .93  &    .24  &    .13 \\
M2   &  1.50  &   2.53  &   1.38  &    .60   &   .71  &    .93  &    .28  &    .17 \\
M3   &  1.55  &   2.93  &   1.55  &    .75   &   .66  &    .92  &    .31  &    .20 \\
M4   &  1.65  &   3.42  &   1.80  &    .83   &   .65  &    .94  &    .40  &    .30 \\
M5   &  1.80  &   4.03  &   2.17  &    .95   &   .62  &    .95  &    .43  &    .35 \\
M6   &  1.95  &   4.65  &   2.55  &   1.06   &   .60  &    .96  &    .46  &    .40 \\
M7   &  2.10  &   5.68  &   3.38  &   1.17   &   .63  &   1.04  &    .53  &    .50 \\
M8   &  ----  &   ----  &   ----  &   ----   &   .74  &   1.24  &    .68  &   ---- \\ \hline
\end{tabular}
\caption{Adopted colors for the dwarfs, in Johnson system.}
\label{tab_phot_dwarfs}
\end{table}

\begin{table}
\tabcolsep 1.89mm
\begin{tabular}{|r|rrrrrrrr|} \hline
Sp Ty   &B-V    &V-I    &V-R    &I-J    &J-H    &J-K    &K-L    &L-M \\
\hline \hline
O5    & -.30    & -.37    & -.15    & -.31    & -.12    & -.19      &.01    & -.05\\
O7    & -.29    & -.36    & -.15    & -.30    & -.11    & -.18      &.02    & -.06\\
O8    & -.27    & -.34    & -.14    & -.28    & -.11    & -.17      &.01    & -.07\\
O9    & -.26    & -.31    & -.12    & -.28    & -.10    & -.17      &.02    & -.02\\
B0    & -.23    & -.27    & -.11    & -.24    & -.09    & -.17    & -.02    & -.01\\
B1    & -.21    & -.25    & -.10    & -.20    & -.09    & -.18    & -.01    & -.01\\
B2    & -.19    & -.24    & -.08    & -.15    & -.08    & -.17    & -.02    & -.02\\
B3    & -.16    & -.18    & -.05    & -.17    & -.05    & -.11    & -.01    & -.02\\
B5    & -.15    & -.18    & -.05    & -.14    & -.03    & -.09      &.01    & -.02\\
B6    & -.13    & -.15    & -.04    & -.13    & -.02    & -.06      &.01    & -.02\\
B7    & -.12    & -.15    & -.04    & -.10    & -.01    & -.04      &.03    & -.02\\
B8    & -.10    & -.10    & -.01    & -.09      &.00    & -.01      &.03    & -.01\\
B9    & -.07    & -.05      &.00    & -.03      &.00    & -.01      &.03      &.00\\
A0    & -.03      &.00      &.03    & -.01      &.02      &.01      &.03      &.00\\
A1      &.01      &.05      &.06      &.01      &.04      &.03      &.03      &.00\\
A2      &.05      &.12      &.09      &.02      &.06      &.05      &.03      &.00\\
A4      &.08      &.22      &.14      &.04      &.10      &.11      &.04      &.00\\
A5      &.15      &.27      &.17      &.06      &.12      &.13      &.04      &.00\\
A7      &.22      &.39      &.21      &.10      &.15      &.17      &.04      &.00\\
A8      &.25      &.44      &.24      &.11      &.17      &.20      &.04      &.00\\
F0      &.30      &.54      &.30      &.14      &.21      &.24      &.05      &.00\\
F2      &.35      &.66      &.35      &.17      &.25      &.28      &.05      &.00\\
F5      &.43      &.81      &.42      &.22      &.31      &.36      &.05      &.00\\
F7      &.50      &.93      &.49      &.24      &.34      &.40      &.06      &.00\\
F8      &.54      &.98      &.52      &.26      &.35      &.43      &.06      &.00\\
G0      &.65     &1.03      &.53      &.28      &.36      &.45      &.07      &.00\\
G2      &.77     &1.10      &.59      &.32      &.40      &.49      &.07      &.00\\
G4      &.83     &1.16      &.65      &.34      &.46      &.55      &.08    & -.01\\
G5      &.86     &1.18      &.67      &.36      &.47      &.56      &.08    & -.01\\
G6      &.89     &1.21      &.70      &.37      &.49      &.58      &.09    & -.02\\
G7      &.91     &1.21      &.73      &.36      &.49      &.58      &.09    & -.02\\
G8      &.94     &1.22      &.76      &.37      &.49      &.58      &.09    & -.01\\
K0     &1.00     &1.29      &.79      &.40      &.52      &.62      &.10    & -.03\\
K1     &1.07     &1.39      &.83      &.44      &.57      &.68      &.11    & -.04\\
K2     &1.16     &1.51      &.87      &.46      &.62      &.73      &.12    & -.05\\
K3     &1.27     &1.75      &.98      &.44      &.66      &.82      &.13    & -.06\\
K4     &1.38     &1.93     &1.05      &.46      &.72      &.88      &.14    & -.07\\
K5     &1.50     &2.03     &1.14      &.64      &.77      &.95      &.15    & -.08\\
K7     &1.53     &2.14     &1.19      &.64      &.81      &.97      &.15    & -.09\\
M0     &1.56     &2.20     &1.22      &.65      &.82     &1.01      &.15    & -.09\\
M1     &1.59     &2.35     &1.27      &.68      &.84     &1.05      &.16    & -.10\\
M2     &1.62     &2.54     &1.37      &.70      &.85     &1.08      &.18    & -.12\\
M3     &----     &2.80     &1.49      &.73      &.88     &1.13      &.20    & -.13\\
M4     &----     &3.20     &1.71      &.86      &.91     &1.17      &.21    & -.14\\
M5     &----     &3.64     &1.97     &----     &----     &1.23     &----     &----\\
M6     &----     &4.28     &2.41     &----     &----     &1.26     &----     &----\\
M7     &----     &5.03     &2.97     &----     &----     &1.27     &----     &----\\
M8     &----     &5.90     &3.61     &----     &----     &----     &----     &----\\ \hline
\end{tabular}
\caption{Adopted colors for the giants, in Johnson system.}
\label{tab_phot_giants}
\end{table}

\begin{table}
\tabcolsep 1.89mm
\begin{tabular}{|r|rrrrrrrr|} \hline
Sp Ty   &B-V    &V-I    &V-R    &I-J    &J-H    &J-K    &K-L    &L-M \\
\hline \hline
O5    & -.32    & -.46    & -.22    & -.25    & -.15    & -.18    & -.03    & -.01\\
O7    & -.31    & -.46    & -.21    & -.23    & -.14    & -.17    & -.03    & -.01\\
O8    & -.30    & -.44    & -.19    & -.24    & -.13    & -.15    & -.03    & -.01\\
O9    & -.27    & -.39    & -.18    & -.21    & -.13    & -.14    & -.03    & -.01\\
B0    & -.22    & -.31    & -.14    & -.17    & -.11    & -.12    & -.01    & -.01\\
B1    & -.19    & -.26    & -.11    & -.15    & -.11    & -.11    & -.01    & -.01\\
B2    & -.16    & -.22    & -.08    & -.12    & -.10    & -.10      &.00      &.00\\
B3    & -.13    & -.17    & -.05    & -.10    & -.09    & -.08      &.00      &.00\\
B5    & -.08    & -.09    & -.01    & -.06    & -.06    & -.06      &.02      &.00\\
B6    & -.06    & -.06      &.00    & -.03    & -.06    & -.06      &.03      &.00\\
B7    & -.04    & -.02      &.02    & -.02    & -.04    & -.02      &.02      &.00\\
B8    & -.03    & -.01      &.02    & -.02    & -.05    & -.02      &.02      &.00\\
B9    & -.01      &.02      &.02      &.01    & -.04    & -.02      &.02      &.00\\
A0      &.00      &.09      &.03    & -.01     &----      &.04      &.05     &----\\
A1      &.02      &.11      &.06    & -.01     &----      &.10      &.05     &----\\
A2      &.04      &.14      &.07    & -.02     &----      &.12      &.05     &----\\
A4      &.09      &.25      &.13    & -.03     &----      &.14      &.05     &----\\
A6      &.12      &.35      &.18      &.01     &----      &.16      &.05     &----\\
F0      &.17      &.41      &.21      &.04     &----      &.19      &.05     &----\\
F2      &.23      &.48      &.27      &.06     &----      &.21      &.05     &----\\
F5      &.32      &.59      &.35      &.08     &----      &.26      &.05     &----\\
F7      &.44      &.65      &.41      &.10     &----      &.32      &.06     &----\\
F8      &.56      &.72      &.45      &.13     &----      &.36      &.06     &----\\
G0      &.76      &.85      &.51      &.19     &----      &.40      &.07     &----\\
G2      &.87      &.99      &.58      &.22     &----      &.47      &.08     &----\\
G4      &.97     &1.07      &.65      &.28     &----      &.50      &.08     &----\\
G5     &1.02     &1.11      &.67      &.32     &----      &.53      &.09     &----\\
G6     &1.06     &1.12      &.67      &.32     &----      &.53      &.10     &----\\
G8     &1.15     &1.15      &.69      &.30     &----      &.54      &.11     &----\\
K0     &1.24     &1.24      &.76      &.34     &----      &.58      &.12     &----\\
K1     &1.30     &1.32      &.80      &.35     &----      &.62      &.13     &----\\
K2     &1.35     &1.40      &.86      &.37     &----      &.66      &.14     &----\\
K3     &1.46     &1.58      &.94      &.42     &----      &.72      &.14     &----\\
K4     &1.53     &1.77     &1.04      &.48     &----      &.76      &.15     &----\\
K5     &1.60     &2.10     &1.21      &.61     &----      &.99      &.15     &----\\
K7     &1.63     &2.14     &1.22      &.63     &----      &.99      &.16     &----\\
M0     &1.63     &2.17     &1.23      &.65     &----      &.97      &.17     &----\\
M1     &1.63     &2.27     &1.27      &.63     &----     &1.02      &.17     &----\\
M2     &1.64     &2.44     &1.33      &.64     &----     &1.03      &.18     &----\\
M3     &1.64     &2.79     &1.47      &.72     &----     &1.07      &.19     &----\\
M4     &1.64     &3.39     &1.73      &.86     &----     &1.02      &.20     &----\\
M5     &1.62     &4.14     &2.17      &.89     &----     &1.02      &.25     &----\\ \hline
\end{tabular}
\caption{Adopted colors for the supergiants, in Johnson system.}
\label{tab_phot_supergiants}
\end{table}
}

\end{document}